\documentstyle[12pt,hpatex]{article}

\begin{document}

\title{The Weinberg Propagators\\}

\authors{Valeri V. Dvoeglazov\footnote{On leave of absence from
{\it Dept. Theor. \& Nucl. Phys., Saratov State University,
Astrakhanskaya ul., 83, Saratov\, RUSSIA.}\,
Internet address: dvoeglazov@main1.jinr.dubna.su}}

\address{Escuela de F\'{\i}sica, Universidad Aut\'onoma de Zacatecas \\
Antonio Doval\'{\i} Jaime\, s/n, Zacatecas 98068,
Zac., M\'exico\\ Internet address:  VALERI@CANTERA.REDUAZ.MX
}

\abstract{An analog of the $j=1/2$ Feynman-Dyson propagator  is
presented in the framework of the $j=1$ Weinberg's theory.
The basis for this construction is the concept of
the Weinberg field as a system of  four field functions
differing  by parity  and  by dual transformations.}

Accordingly  to the Feynman-Dyson-Stueckelberg ideas,
a causal propagator  has to be constructed
by using  the formula (e.~g., ref.~\cite[p.91]{Itzykson})
\begin{eqnarray}
\lefteqn{S_F (x_2, x_1) =\int \frac{d^3 k}{(2\pi)^3}\frac{m}{E_k} \left [
\theta (t_2 -t_1) \, a \,\,
u^\sigma (k) \otimes \overline u^{\sigma} (k) e^{-ikx} + \right. }\nonumber\\
&&\left. \qquad\qquad  +\, \,\theta (t_1 - t_2) \, b \,\,
v^\sigma (k) \otimes \overline v^\sigma (k) e^{ikx} \right ] \quad ,
\end{eqnarray}
$x=x_2 -x_1$. In the $j=1/2$ Dirac theory  it results  to
\begin{equation}\label{dp}
S_F (x) = \int \frac{d^4 k}{(2\pi)^4} e^{-ikx} \frac{\hat k +m}{k^2 -m^2
+i\epsilon} \quad,
\end{equation}
provided that the constant $a$ and $b$ are determined by imposing
\begin{equation}
(i\hat \partial_2 -m) S_F (x_2, x_1) =\delta^{(4)} (x_2 -x_1) \quad,
\end{equation}
namely, $a=-b=1/ i$ .

However,  attempts to construct the covariant propagator in this way
have failed in the framework of the Weinberg theory,
ref.~\cite{Weinberg}, which is a generalization of the Dirac's ideas to
higher spins. For instance, on the page B1324 of ref.~\cite{Weinberg}\,
Weinberg writes:

\smallskip

``{\it Unfortunately,
the propagator arising from Wick's theorem is  NOT equal to the covariant
propagator
except for $j=0$ and $j=1/2$. The trouble is that the derivatives act on the
$\epsilon (x) = \theta (x) - \theta (-x)$ in $\Delta^C (x)$ as well as on
the functions\footnote{In the cited paper $\Delta_1(x) \equiv i \left [\Delta_+
(x) + \Delta_+ (-x)\right ]$ and $\Delta (x) \equiv \Delta_+ (x) - \Delta_+
(-x)$ have been used.
 $i\Delta_+ (x) \equiv \frac{1}{(2\pi)^3} \int \frac{d^3 p}{2E_p} \exp
(ipx)$ is a particle Green's function.} $\Delta$ and $\Delta_1$. This
gives rise to extra terms proportional to equal-time $\delta$ functions
and their derivatives\ldots The cure is well known: \ldots compute the
vertex factors using only the original covariant part of the Hamiltonian
${\cal H}$; do not use the Wick propagator for internal lines; instead use
the covariant propagator.}

\smallskip

The propagator, recently proposed  in refs.~\cite{Ahl2,Ah-pro}
(see also ref.~[3]), is the causal propagator.
However, the old problem remains: the Feynman-Dyson propagator
is not the Green's function of the Weinberg equation. As mentioned,
the covariant propagator proposed by Weinberg propagates kinematically
spurious solutions~\cite{Ah-pro}\ldots
The aim of my paper is to consider
the problem of constructing the propagator in
the framework of the model given in~\cite{D1,D2}.
The concept of
the Weinberg field ``doubles"  has been proposed there.
It is based on the equivalence between
a Weinberg field and an  antisymmetric tensor field, ref.~\cite{D1},
which can be described by $F_{\mu\nu}$ and
its dual $\tilde F_{\mu\nu}$.
These field functions may be used to form a parity doublet. An essential
ingredient of the consideration of ref.~\cite{D2}
is the idea of combining the Lorentz and the dual
transformation. This idea, in fact, has been proposed
in refs.~\cite{Ahl,Ahl2}. An example of such combining is
a Bargmann-Wightman-Wigner-type quantum field theory,~ref.~[3b].

The set of four equations has been proposed in ref.~\cite{D1}.
For the functions $\psi_1^{(1)}$ and $\psi_2^{(1)}$, connected with the
first one by  the dual (chiral, $\gamma_5$)  transformation,
the equations are
\begin{eqnarray}\label{eq:a1}
(\gamma_{\mu\nu} p_\mu p_\nu +m^2 )\psi_1^{(1)} &=& 0\quad,\\ \label{eq:a2}
(\gamma_{\mu\nu} p_\mu p_\nu - m^2) \psi_2^{(1)} &=& 0 \quad.
\end{eqnarray}
For the field functions
connected with $\psi_1^{(1)}$ and $\psi_2^{(1)}$ by $\gamma_5\gamma_{44}$
transformations the set of equations is written:
\begin{eqnarray}\label{eq:a11}
\left [\widetilde \gamma_{\mu\nu}p_\mu p_\nu - m^2\right ] \psi_1^{(2)} &=&0
\quad,\\
\label{eq:a21}
\left [\widetilde \gamma_{\mu\nu} p_\mu p_\nu + m^2 \right ] \psi_2^{(2)} &=&0
\quad,
\end{eqnarray}
where $\widetilde \gamma_{\mu\nu} = \gamma_{44} \gamma_{\mu\nu} \gamma_{44}$
is connected with the Barut-Muzinich-Williams $j=1$ matrices~\cite{Barut}.

In the cited papers  I  have used the plane-wave expansion
\begin{eqnarray}\label{pl1}
\psi_1 (x) &=&\sum_\sigma \int \frac{d^3 p}{(2\pi)^3} \frac{1}{m \sqrt{2E_p}}
\left [ u_1^\sigma (\vec p) a_\sigma (\vec p) e^{ipx} +v_1^\sigma (\vec p)
b^\dagger_\sigma (\vec p) e^{-ipx} \right ]\quad,\\
\label{pl2}
\psi_2 (x) &=&\sum_\sigma \int \frac{d^3 p}{(2\pi)^3} \frac{1}{m\sqrt{2E_p}}
\left [ u_2^\sigma (\vec p) c_\sigma (\vec p) e^{ipx} +v_2^\sigma (\vec p)
d^\dagger_\sigma (\vec p) e^{-ipx} \right ] \quad,
\end{eqnarray}
where $E_p=\sqrt{\vec p^{\,2} +m^2}$,  in order
to prove that  one can describe a
$j=1$ quantum particle with transversal components in the framework
of  the Weinberg and/or  of  the antisymmetric tensor theory.

The corresponding bispinors
in the momentum space coincide with the Tucker-Hammer ones within
a normalization.\footnote{They  also coincide with the bispinors of Ahluwalia
{\it et al.}, ref.~\cite{Ahl2}, within a unitary transformation.} Their
explicit form is
\begin{eqnarray}\label{b1}
u_1^{\sigma\, (1)} (\vec p)= v_1^{\sigma\, (1)}
(\vec p) =\frac{1}{\sqrt{2}}\pmatrix{\left [m+ (\vec J\vec p)
+{(\vec J \vec p)^2 \over  (E+m)}\right ]\xi_\sigma \cr \left [ m  -
(\vec J\vec p) +{(\vec J \vec p)^2 \over  (E+m)}\right ]
\xi_\sigma} \quad,
\end{eqnarray}
and
\begin{eqnarray}\label{b2}
u_2^{\sigma\,(1)} (\vec p)= v_2^{\sigma\,(1)} (\vec p)
=\frac{1}{\sqrt{2}}\pmatrix{\left [m+
(\vec J\vec p) +{(\vec J \vec p)^2 \over  (E+m)}\right ]\xi_\sigma \cr
\left [ - m  + (\vec J\vec p) - {(\vec J \vec p)^2
\over  (E+ m)}\right ] \xi_\sigma}\quad.
\end{eqnarray}
Thus,  $u_2^{(1)} (\vec p) = \gamma_5 u_1^{(1)} (\vec p)$ and
$\overline u_2^{(1)} (\vec p) = -\overline u_1^{(1)} (\vec p)\gamma_5$.

Bispinors
\begin{eqnarray}\label{b11}
u_1^{\sigma\, (2)} (\vec p)= v_1^{\sigma\, (2)}
(\vec p) =\frac{1}{\sqrt{2}}\pmatrix{\left [m- (\vec J\vec p)
+{(\vec J \vec p)^2 \over  (E+m)}\right ]\xi_\sigma \cr \left [ -m  -
(\vec J\vec p) -{(\vec J \vec p)^2 \over  (E+m)}\right ]
\xi_\sigma}\quad,
\end{eqnarray}
\begin{eqnarray}\label{b21}
u_2^{\sigma\,(2)} (\vec p)= v_2^{\sigma\,(2)} (\vec p)
=\frac{1}{\sqrt{2}}\pmatrix{\left [-m+
(\vec J\vec p) -{(\vec J \vec p)^2 \over  (E+m)}\right ]\xi_\sigma \cr
\left [  -m  - (\vec J\vec p) - {(\vec J \vec p)^2
\over  (E+ m)}\right ] \xi_\sigma}
\end{eqnarray}
satisfy Eqs. (\ref{eq:a11}) and (\ref{eq:a21}) written in the momentum space.
Thus,
$u_1^{(2)} (\vec p) = \gamma_5\gamma_{44} u_1^{(1)} (\vec p)$,
$\overline u_1^{(2)} = \overline u_1^{(1)} \gamma_5\gamma_{44}$,
$u_2^{(2)} (\vec p) = \gamma_5\gamma_{44} \gamma_5 u_1^{(1)} (\vec p)$
and $\overline u_2^{(2)} (\vec p) = - \overline u_1^{(1)}\gamma_{44}$.

Let me check, if the sum of four equations
($x=x_2 -x_1$)
\begin{eqnarray}
&&\hspace*{-1cm}\left [ \gamma_{\mu\nu} \partial_\mu \partial_\nu -m^2 \right ]
 \int  \frac{d^3 p}{(2\pi)^3 2E_p}
\left [ \theta (t_2 -t_1) \, a\,\,\,   u_1^{\sigma\,(1)} (p) \otimes \overline
u_1^{\sigma\,(1)} (p) e^{ipx}+\right .\nonumber\\
&&\left.  \qquad\qquad+\theta (t_1 -t_2) \, b \,\,\, v_1^{\sigma\,(1)} (p)
\otimes \overline  v_1^{\sigma\,(1)} (p) e^{-ipx} \right  ] +\nonumber\\
&+& \left [ \gamma_{\mu\nu} \partial_\mu \partial_\nu + m^2 \right  ]  \int
\frac{d^3 p}{(2\pi)^3 2E_p}
\left [ \theta (t_2 -t_1) \, a\,\,\,   u_2^{\sigma\,(1)} (p) \otimes \overline
u_2^{\sigma\,(1)} (p) e^{ipx}+
\right. \nonumber\\
&&\left. \qquad\qquad+\theta (t_1 -t_2) \, b \,\,\, v_2^{\sigma\,(1)} (p)
\otimes \overline  v_2^{\sigma\,(1)} (p) e^{-ipx}\right  ] +\nonumber\\
&+&\left [ \widetilde \gamma_{\mu\nu} \partial_\mu \partial_\nu + m^2 \right  ]
\int \frac{d^3 p}{(2\pi)^3 2E_p}
\left [ \theta (t_2 -t_1) \, a\,\,\,  u_1^{\sigma\,(2)} (p) \otimes \overline
u_1^{\sigma\,(2)} (p) e^{ipx}+ \right.\nonumber\\
&&\left. \qquad\qquad+\theta (t_1 -t_2) \, b \,\,\, v_1^{\sigma\,(2)} (p)
\otimes \overline  v_1^{\sigma\,(2)} (p)e^{-ipx} \right ] +\nonumber\\
&+&\left [\widetilde \gamma_{\mu\nu} \partial_\mu \partial_\nu - m^2 \right  ] \int
\frac{d^3 p}{(2\pi)^3 2E_p}
\left [ \theta (t_2 -t_1) \, a\,\,\,   u_2^{\sigma\,(2)} (p) \otimes \overline
u_2^{\sigma\,(2)} (p)  e^{ipx} +\right.\nonumber\\
&&\left.  \qquad\qquad+\theta (t_1 -t_2) \, b \,\,\, v_2^{\sigma\,(2)} (p)
\otimes \overline  v_2^{\sigma\,(2)} (p) e^{-ipx} \right ] =
\delta^{(4)} (x_2 -x_1)
\end{eqnarray}
can be satisfied by the definite choice of $a$ and $b$.
The relation  $ u_i (p) =  v_i (p)$ for bispinors in the momentum space
had been also used in ref.~\cite{D2,Hammer}.  In the process of
calculations  I  assume
that the set of``Pauli spinors"\footnote{I mean their analogues in the
$(1,0)$ or $(0,1)$
spaces.}  is the complete set and it is normalized
to $\delta_{\sigma\sigma^\prime}$\, .

The simple calculations yield
\begin{eqnarray}
\lefteqn{\partial_\mu \partial_\nu  \left [ a\, \theta (t_2 -t_1)\, e^{ip(x_2
-x_1)} + b\, \theta (t_1 -t_2)\, e^{-ip(x_2 -x_1)} \right ]=\nonumber}\\
&=& - \left [ a\, p_\mu p_\nu \theta (t_2 - t_1)
\exp \left [ ip(x_2 -x_1)\right ] +
b\,  p_\mu p_\nu  \theta (t_1 -t_2)
\exp \left [ -ip (x_2 -x_1) \right ] \right
] + \nonumber\\
&+& a\left [ - \delta_{\mu 4} \delta_{\nu 4} \delta^{\,\,\prime}
(t_2 -t_1) +i (p_\mu \delta_{\nu 4} +p_\nu \delta_{\mu 4}) \delta (t_2 -t_1)
\right ] \exp \left [i \vec p
(\vec x_2 - \vec x_1)\right ] +\nonumber\\
&+& b\, \left [ \delta_{\mu 4} \delta_{\nu 4} \delta^{\,\,\prime}
(t_2 -t_1) + i (p_\mu \delta_{\nu 4} +p_\nu \delta_{\mu 4})
\delta (t_2 -t_1) \right ] \exp \left [-i\vec p
(\vec x_2 - \vec x_1)\right ] \quad;
\end{eqnarray}
and
\begin{eqnarray}
u_1^{(1)}\overline u_1^{(1)} ={1\over 2} \pmatrix{m^2 & S_p \otimes S_p\cr
\overline S_p \otimes \overline S_p &m^2\cr}\quad,\quad
u_2^{(1)}\overline u_2^{(1)} = {1\over 2}\pmatrix{-m^2 & S_p \otimes S_p\cr
\overline S_p \otimes \overline S_p &-m^2\cr}\quad,
\end{eqnarray}
\begin{eqnarray}
u_1^{(2)}\overline u_1^{(2)} ={1\over 2} \pmatrix{-m^2 &
\overline S_p \otimes \overline
 S_p\cr S_p \otimes  S_p &-m^2\cr}\quad,\quad
 u_2^{(2)}\overline u_2^{(2)} = {1\over 2}
\pmatrix{m^2 & \overline S_p \otimes \overline S_p\cr S_p
\otimes  S_p &m^2\cr}\quad,
\end{eqnarray}
where
\begin{eqnarray}
S_p &=& m + (\vec J \vec p) +\frac{(\vec J \vec p)^2}{E+m}\quad,\\
\overline S_p &=& m - (\vec J \vec p) + \frac{(\vec J \vec p)^2}{E+m}\quad.
\end{eqnarray}
Due to
$$\left [E_p - (\vec J \vec p)\right ]  S_p \otimes S_p = m^2 \left [ E_p
+ (\vec J \vec p)\right ]\quad,$$
$$\left [E_p + (\vec J \vec p)\right ] \overline S_p
\otimes \overline S_p = m^2 \left [ E_p - (\vec J \vec p)\right ]\quad.$$
one  can conclude: the  generalization of
the notion of causal  propagators is  admitted by using
``Wick's formula" for the time-ordered particle operators
provided that  $a=b=1/ 4im^2$. It is necessary to  consider
all four equations, Eqs. (\ref{eq:a1})-(\ref{eq:a21}).

The $j=1$ analogues of the formula (\ref{dp})  for the Weinberg propagators
follow from the formula (3.6) of ref.~\cite{Ahl2}
immediately:\footnote{Please do not forget that I use
the Euclidean metric as in my previous papers.}
\begin{equation}\label{propa1}
S_F^{(1)} ( p ) \sim -\frac{1}{i(2\pi)^4  (p^2  +m^2 -i\epsilon)} \left [
\gamma_{\mu\nu} p_\mu p_\nu   -  m^2  \right ]\quad,
\end{equation}
\begin{equation}\label{propa2}
S_F^{(2)} ( p ) \sim -\frac{1}{i(2\pi)^4  (p^2  +m^2 -i\epsilon)} \left [
\gamma_{\mu\nu} p_\mu p_\nu   +  m^2  \right ]\quad,
\end{equation}
\begin{equation}\label{propa3}
S_F^{(3)} ( p ) \sim -\frac{1}{i(2\pi)^4  (p^2  +m^2 -i\epsilon)} \left [
\widetilde\gamma_{\mu\nu} p_\mu p_\nu   +  m^2  \right ] \quad,
\end{equation}
\begin{equation}\label{propa4}
S_F^{(4)} ( p ) \sim -\frac{1}{i(2\pi)^4  (p^2  +m^2 -i\epsilon)} \left [
\widetilde \gamma_{\mu\nu} p_\mu p_\nu   -  m^2  \right ] \quad.
\end{equation}

It is interesting to note that the causal
propagator consisting of four terms, four parts, four
propagators has been met earlier. Namely,
in the bound state theory. You may compare
the propagators which is above with the Green's
function for the two-fermion system,
ref.~\cite{Bound,Tyukh}:\footnote{For a recent review see
ref.~\cite{Our}.}
\begin{eqnarray}
G_0 &=& i (2\pi)^4 \delta (p-q) S_1 (p_1) S_2 (p_2)\quad,\\
S_i &=& - \left [ \Lambda_i^+ (\vec p_i) (p_{0i} - E_{ip} +i\epsilon)^{-1}
+\Lambda_i^- (\vec p_i)
(p_{0i} + E_{ip} +i\epsilon)^{-1}\right ] \gamma_{i0}\quad,
\end{eqnarray}
$\Lambda_i^{\pm}$ are the projection operators.

We should use the obtained set of  Weinberg propagators
(\ref{propa1},\ref{propa2},\ref{propa3},\ref{propa4})
in perturbation calculus of scattering amplitudes.
In ref.~\cite{Dvoegl} the amplitude for the interaction
of two $2(2j+1)$ bosons has been obtained on the basis
of the use of one field only and it is obviously incomplete,
see also ref.~\cite{Hammer}. But, it is interesting
that the spin structure has proved there
not to be changed  regardless we consider
the two-Dirac-fermion interaction or the two-Weinberg($j=1$)-boson
interaction. However, the denominator slightly differs ($1/\vec \Delta^2
\rightarrow 1/2m(\Delta_0 -m)$) in the cited papers~\cite{Dvoegl}
from the fermion-fermion case. More accurate consideration
of the fermion-boson and boson-boson interactions in the framework
of the Weinberg theory is in progress.

\smallskip

The conclusions are: one can construct an analog of the Feynman-Dyson
propagator for the $2(2j+1)$ model and, hence, a ``local"
theory provided that the Weinberg states are
``quadrupled"  ($j=1$ case).

\smallskip

{\bf Acknowledgments.}
The papers of  Dr.~D. V. Ahluwalia
led me to the ideas presented here and
in my previous articles. His answers on my questions were very helpful
in realizing the importance of presented topics.
I thank  him, Dr.  I.~G.~Kaplan and  Dr. A.~Mondragon  for  encouragements;
Dr.~Yu.~F.~Smirnov for asking the right questions at the right time.

I am grateful to Zacatecas University for professorship.

\end{document}